\documentclass{article}
\usepackage{graphicx}
\usepackage{verbatim}
\usepackage{latexsym}
\usepackage{amssymb}
\usepackage{amsmath}
\usepackage{amsthm}
\usepackage[english]{babel}
\usepackage{babel}
\newcommand{\F}{\mathcal{F}}

\newtheorem{theorem}{Theorem}
\newtheorem{lemma}{Lemma}
\newtheorem{corollary}{Corollary}
\newtheorem{proposition}{Proposition}

\newcommand{\Rel}{\mathcal{R}}

\newcommand{\compR}{\overline{\mathcal{R}}}
\newcommand{\LR}{\mathcal{L}(\Rel)}

\newcommand{\bipR}{bip(\mathcal{R})}

\begin{document}

\title{The Dilworth Number of Auto-Chordal-Bipartite Graphs}


\author{
Anne Berry
\footnote{LIMOS UMR CNRS 6158, Universit\'e Blaise
Pascal, F-63 173 Aubi\`ere, France \texttt{berry@isima.fr}}
\and Andreas Brandst\"{a}dt
\footnote{Institut f\"ur Informatik, Universit\"at
Rostock, D-18051 Rostock, Germany
\texttt{andreas.brandstaedt@uni-rostock.de}}
\and Konrad Engel
\footnote{Institut f\"ur Mathematik, Universit\"at
Rostock, D-18051 Rostock, Germany
\texttt{konrad.engel@uni-rostock.de}}
}


\maketitle
\begin{abstract}

The {\em mirror} (or {\em bipartite complement}) $mir(B)$ of a bipartite graph $B=(X,Y,E)$ has the same color classes $X$ and $Y$ as $B$, and two
vertices $x \in X$ and $y \in Y$ are adjacent in $mir(B)$ if and only if $xy \notin E$. A bipartite graph is {\em chordal bipartite} if none of its induced
subgraphs is a chordless cycle with at least six vertices. In this paper, we deal with chordal bipartite graphs whose mirror is chordal bipartite as well;
we call these graphs {\em auto-chordal bipartite graphs} ({\em ACB graphs} for short). We describe the relationship to some known graph classes such as interval and strongly chordal graphs and we present several characterizations of ACB graphs. We show that ACB graphs have unbounded Dilworth number, and we characterize ACB graphs with Dilworth number $k$.


\end{abstract}


\section{Introduction}\label{Introduction}
\par
Given a finite relation $\Rel$ between two sets $X$ and $Y$, a corresponding graph can classically be defined in several ways;
$X$ and $Y$ are often considered either as stable sets or as cliques of a graph, and $\Rel$ describes the edges between $X$ and $Y$.
First, when both $X$ and $Y$ are stable sets, $\Rel$ defines a \emph{bipartite graph} $\bipR$ with edges between $X$ and $Y$.
The maximal bicliques of this bipartite graph can be organized by inclusion into a lattice, called a \textit{concept lattice} $\LR$ \cite{GanWil1999}
(or \textit{Galois lattice} \cite{CasLecMon2007}).
Second, when both $X$ and $Y$ are cliques, the corresponding graph is \textit{co-bipartite}.
Third, when without loss of generality, $X$ is a clique and $Y$ is a stable set, the corresponding graph is a \emph{split graph}.
Finally, $\Rel$ defines a \textit{hypergraph} where, without loss of generality, $X$ is the vertex set and $Y$ describes the hyperedges.
\par
Naturally, there are strong relationships between different realizations of $\Rel$.
One example of this correspondence, which is central to this paper, is the one between chordal bipartite graphs
and strongly chordal graphs (see \cite{Dahlh1991} and Lemma \ref{lemma:cbgstrchsplit} of Section \ref{Sect:Preliminaries}):
A bipartite graph $B=(X,Y,E)$ is chordal bipartite if and only if the graph obtained from $B$ by completing $X$ to a clique is strongly chordal.
\par
In this paper, we will also use the complement relation $\compR$, which we called the \textit{mirror relation} \cite{BerSig2012}.
The {\em mirror} (or {\em bipartite complement}) $mir(B)$ of a bipartite graph $B=(X,Y,E)$ has the same color classes $X$ and $Y$ as $B$, and two
vertices $x \in X$ and $y \in Y$ are adjacent in $mir(B)$ if and only if $xy \notin E$.
Several papers use this mirror notion, with various names and notations.
Most of them in fact investigate '\textit{auto-mirror}' relations (\textit{i.e.}, both the relation and its mirror relation are in the same class).
Such relations were used e.g. by \cite{FouGiaVan1999} to describe bipartite graphs whose vertex set can be partitioned into a stable set and a maximal biclique;
by \cite{GiaVan2003} to decompose a bipartite graph in a manner similar to modular decomposition;
by \cite{MaSpi1994} to investigate the chain dimension of a bipartite graph, remarking the (obvious) fact that a bipartite graph is a chain graph (i.e., is $2K_2$-free) if and only if its mirror is also a chain graph; by \cite{BenHamDeW1985} and \cite{FoeHam1977/2} to characterize split graphs of Dilworth number 2 (the \emph{Dilworth number} of a graph is the maximum number of its vertices such that the neighborhood of one vertex is not contained in the closed neighborhood of another \cite{Spinr2003}; see Section \ref{Sect:Dilworth}); by \cite{Nara1982} to characterize split graphs of Dilworth number 3;
by \cite{BerSig2012} to characterize lattices with an articulation point.
\par
Recently, \cite{BerSig2013} characterized concept lattices which are planar and whose mirror lattice is also planar: this is the case if and only if the
corresponding bipartite graph as well as its mirror is chordal bipartite.
We call these graphs \textit{auto-chordal-bipartite graphs} (\textit{ACB graphs} for short); these are the main topic of this paper.
Though chordal bipartite graphs have given rise to a wealth of publications, to the best of our knowledge ACB graphs have not been studied.
\par
By Lemma \ref{lemma:cbgstrchsplit}, ACB graphs correspond to split graphs which are strongly chordal and whose mirror is strongly chordal as well (\textit{auto-strongly-chordal} graphs). One special class of auto-strongly-chordal graphs which is well-known is that of interval graphs whose complement is an interval graph as well (\textit{auto-interval graphs}); this special class of split graphs was characterized by \cite{BenHamDeW1985} using results from \cite{FoeHam1977/2} as those having Dilworth number at most 2.
\par
This paper is organized as follows: In Sections \ref{Sect:Preliminaries} and \ref{Sect:Dilworth}, we give some necessary notations, definitions and previous results.
In Section \ref{Sect:DilwACB-XY}, we show that the Dilworth number of ACB graphs is unbounded. We address the question of determining both the Dilworth number with respect to $X$ and to $Y$, and show that both numbers can be arbitrarily large and that the gap between the two numbers can also be arbitrarily large.
In Section \ref{Sect:DilwACBbounded}, the main result of this paper is a characterization of ACB graphs with Dilworth number at most $k$ in terms of forbidden induced subgraphs.
Finally, in Section \ref{Sect:Algo}, we discuss some algorithmic aspects of ACB graphs.

\section{Notions and Preliminary Results}\label{Sect:Preliminaries}
\subsection{Some Basic Graph and Hypergraph Notions}\label{Subsect:GraphNotions}

Throughout this paper, all graphs are finite, simple (i.e., without loops and multiple edges) and undirected. For a graph $G=(V,E)$, let
$\overline{G}=(V,\overline{E})$ denote the complement graph with $\overline{E}=\{xy: x \neq y, xy \notin E\}$. Isomorphism of graphs $G_1$, $G_2$ will be denoted by $G_1 \sim G_2$.
As usual, $N(x)=\{y \in V: xy \in E\}$ is the \textit{open neighborhood} of $x$, and $N[x]=N(x) \cup \{x\}$ is the \textit{closed neighborhood} of $x$

For $X \subseteq V$, $G[X]$ denotes the subgraph induced by $X$. For a set ${\cal F}$ of graphs, $G$ is {\em ${\cal F}$-free} if none of the induced subgraphs of $G$ is in ${\cal F}$. A \textit{clique} is a set of pairwise adjacent vertices. A \textit{stable set} or \textit{independent set} is a set of pairwise non-adjacent vertices.

For two vertex-disjoint graphs $G_1$ and $G_2$, $G_1+G_2$ denotes the disjoint union of them; $iK_2$ denotes the disjoint union of $i$ edges, $i \ge 2$.

\par

$C_k$, $k\geq 4$, denotes the chordless cycle on $k$ vertices. A graph is {\em chordal} if it is $C_k$-free for every $k \ge 4$.
$P_k$, $k\geq 4$, denotes the chordless path on $k$ vertices. For $k \ge 3$, a (complete) {\em $k$-sun}, denoted $S_k$, consists of a clique with $k$ vertices, say $q_0,\ldots,q_{k-1}$, and another $k$ vertices, say $s_0,\ldots,s_{k-1}$, such that $s_0,\ldots,s_{k-1}$ form a stable set and every $s_i$ is adjacent to exactly $q_i$ and $q_{i+1}$ (index arithmetic modulo $k$). Later on, $S_3$, $S_4$ and $\overline{S_3}$ (also called {\em net}) play a special role. A chordal graph is {\em strongly chordal} \cite{Farbe1983} if it is $S_k$-free for every $k \ge 3$.

\par

A \textit{bipartite graph} $B$ is a graph whose vertex set can be partitioned into two stable sets $X$ and $Y$, which we refer to as its \textit{color classes}. We use the notation $B=(X,Y,E)$. A \textit{biclique} in $B=(X,Y,E)$ is a subgraph induced by sets $X' \subseteq X$ and $Y' \subseteq Y$ having all possible edges between elements of $X'$ and $Y'$. In a bipartite graph $B=(X,Y,E)$, vertices of a path $P_k$ alternate between $X$ and $Y$;
a $P_7$ with its end-vertices in $X$ (with its end-vertices in $Y$, respectively) is called an {\em $X$-$P_7$} (a {\em $Y$-$P_7$}, respectively).

A bipartite graph $B$ is a \textit{chordal bipartite graph} if $B$ is $C_{2k}$-free for all $k \ge 3$ \cite{GolGos1978}.
A \textit{chain graph} is a $2K_2$-free bipartite graph; obviously, every chain graph is chordal bipartite.
A \textit{co-bipartite} graph is the complement of a bipartite graph, i.e., a graph whose vertex set can be partitioned into two cliques $X$ and $Y$.
A graph $G$ is a \textit{split graph} if its vertex set can be partitioned into a clique $K$ and a stable set $I$, also denoted as $G=(K,I,E)$.
The following is well-known:

\begin{lemma}[\cite{FoeHam1977/1}]\label{lemma:splitgrchar}
The following are equivalent:
\begin{enumerate}
\item[$(i)$] $G$ is a split graph.
\item[$(ii)$] $G$ and $\overline{G}$ are chordal.
\item[$(iii)$] $G$ is $(2K_2,C_4,C_5)$-free.
\end{enumerate}
\end{lemma}

For a given bipartite graph $B=(X,Y,E)$, let $split_X(B)$ ($split_Y(B)$, respectively) denote the split graph resulting from $B$ by completing $X$ ($Y$, respectively) to a clique. For example, if $B=C_{2k}$ then $split_X(B)=S_k$ and if $B=3K_2$ then $split_X(B)=\overline{S_3}$, $k \ge 3$.

The following fact is well-known (\cite{Dahlh1991}, see also \cite{BraLeSpi1999}):

\begin{lemma}[\cite{Dahlh1991}]\label{lemma:cbgstrchsplit}
A bipartite graph $B=(X,Y,E)$ is chordal bipartite if and only if $split_X(B)$ $(split_Y(B)$, respectively$)$ is strongly chordal.
\end{lemma}

For a split graph $G=(K,I,E)$ with split partition into a clique $K$ and a stable set $I$, let $bip(G)=(K,I,E')$ denote the corresponding bipartite graph with edge set $E'=\{xy: x \in K, y \in I, xy \in E\}$.

\par

The reader is referred to \cite{BraLeSpi1999} and  \cite{Spinr2003} for further graph notions.

\subsection{Mirror of Relations, Hypergraphs, Bipartite Graphs and Split Graphs}\label{Subsect:MirrorNotions}

The notion of mirror of relations, hypergraphs, bipartite graphs and split graphs is closely related to the complement and is defined as follows:

\begin{enumerate}
\item[$(1)$] Let $\Rel \subseteq X \times Y$ be a relation between sets $X$ and $Y$. The \textit{mirror relation} of $\Rel$, denoted $mir(\Rel)=\compR$,  is the complement relation $\compR \subseteq X \times Y$ such that $(x,y) \in \compR$  if and only if $(x,y) \not\in\Rel$.

\item[$(2)$] Let $H=(V,{\cal E})$ be a hypergraph. The {\em mirror} of $H$ is the complement hypergraph
$mir(H)=\overline{H}=(V,\{\overline{e}: e \in {\cal E}\})$.

\item[$(3)$] Let $B=(X,Y,E)$ be a bipartite graph. The \textit{mirror} (or {\em bipartite complement}) of  $B$ is
the bipartite graph $mir(B)=(X,Y,E')$ such that for all $x \in X$, $y \in Y$, $xy \in E'$ if and only if
$xy\not\in E$. Thus, for example, $mir(C_6)=3K_2$, $mir(C_8)=C_8$, and $mir(P_7)=P_7$.

\item[$(4)$] Let $G=(K,I,E)$ be a split graph with split partition into a clique $K$ and stable set $I$. The {\em mirror of $G$} is the split graph $mir(G) = (K,I,E')$ where for all $x \in K$ and for all $y \in I$, $xy \in E'$ if and only if $xy\not\in E$.

\end{enumerate}

\begin{figure}
\begin{center}
\begin{tabular}{|c|c|c|c|}
\hline
bipartite $B$ & $mir(B)$ & $Split_X(B)$  & $mir(Split_X(B))$ \\
\hline
 ~ &~ &~ &~ \\
\includegraphics[height=1cm]{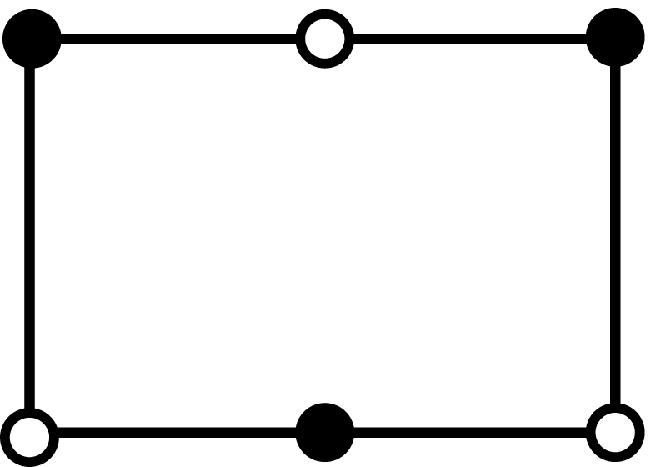}& \includegraphics[height=1cm]{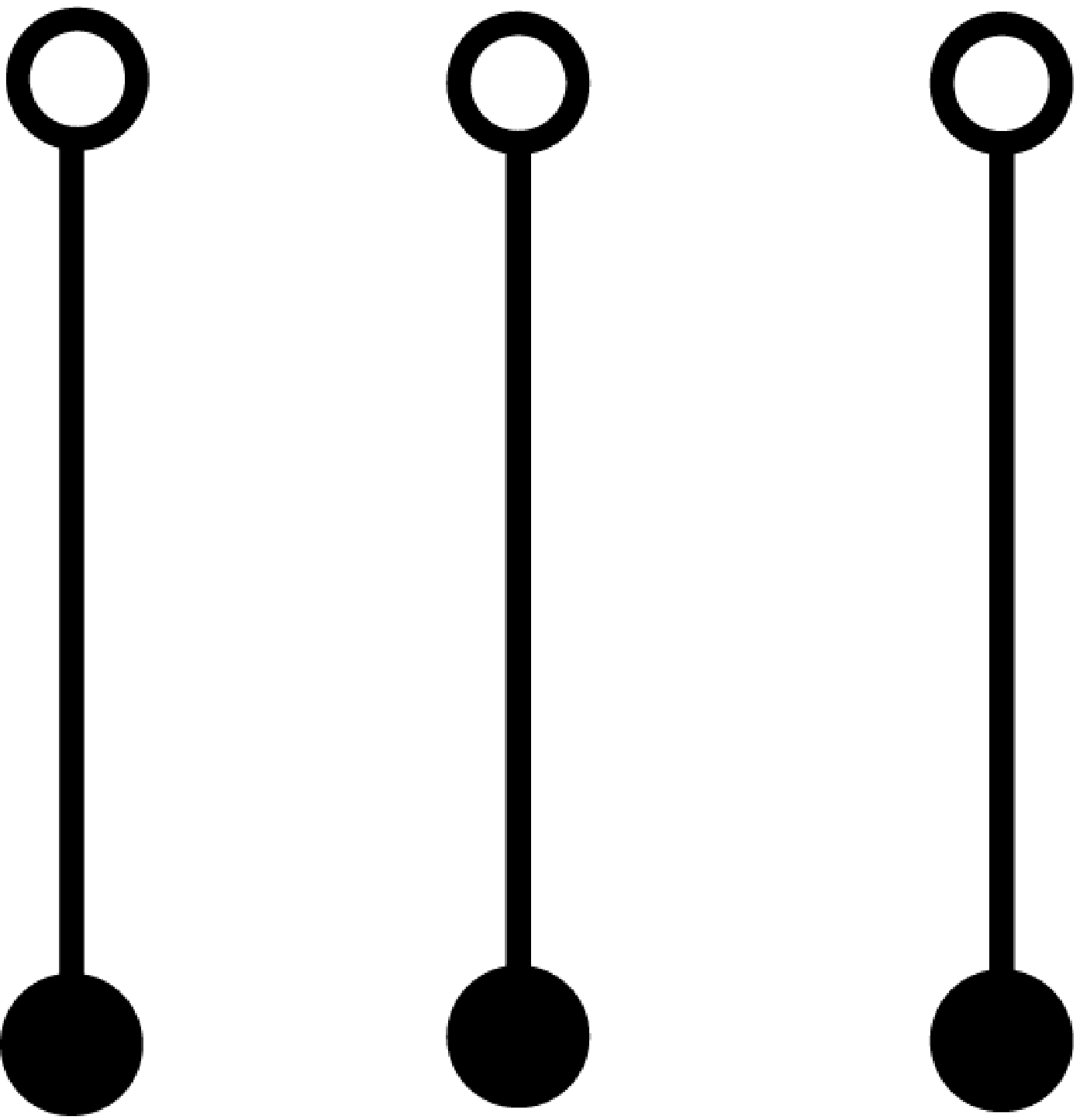} &  \includegraphics[height=1.1cm]{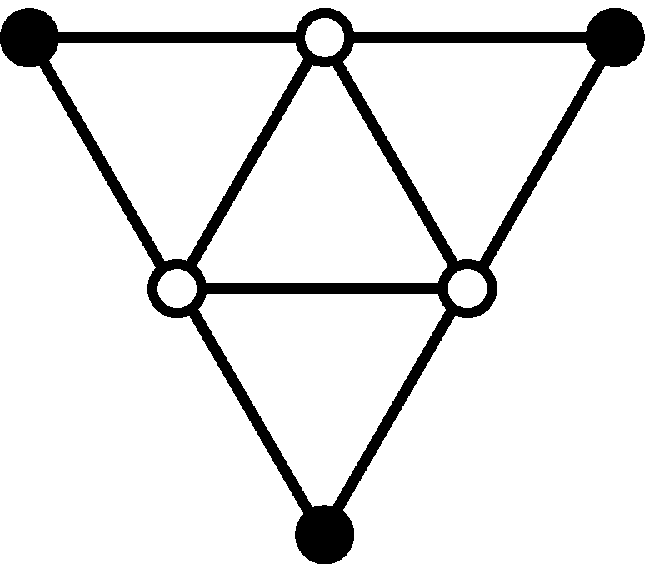}&  \includegraphics[height=1.4cm]{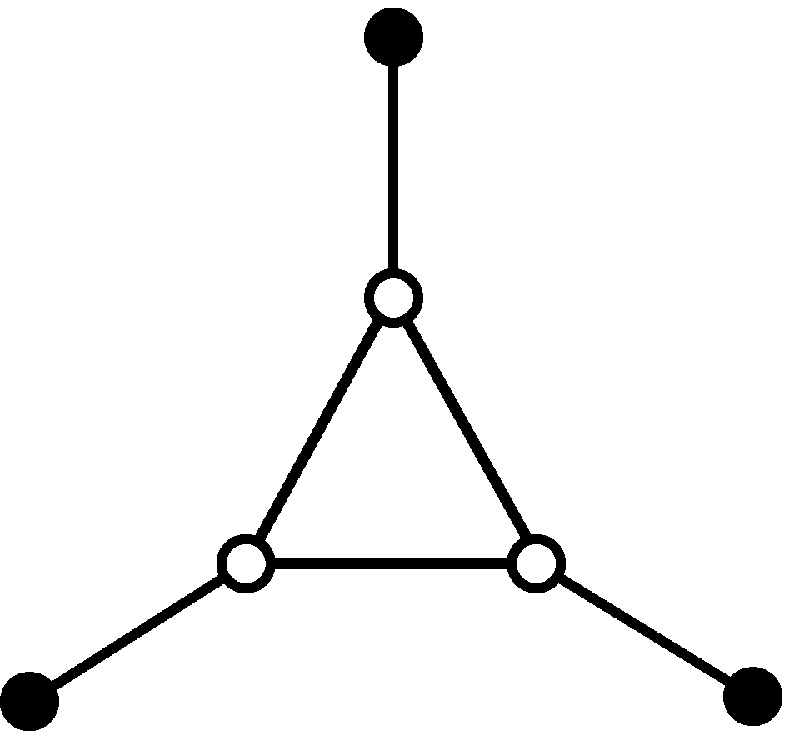}\\

$C_6$& $3K_2$  &$S_3$ & $\overline{S_3}$\\
\hline
 ~ &~ &~ &~\\
\includegraphics[height=.9cm]{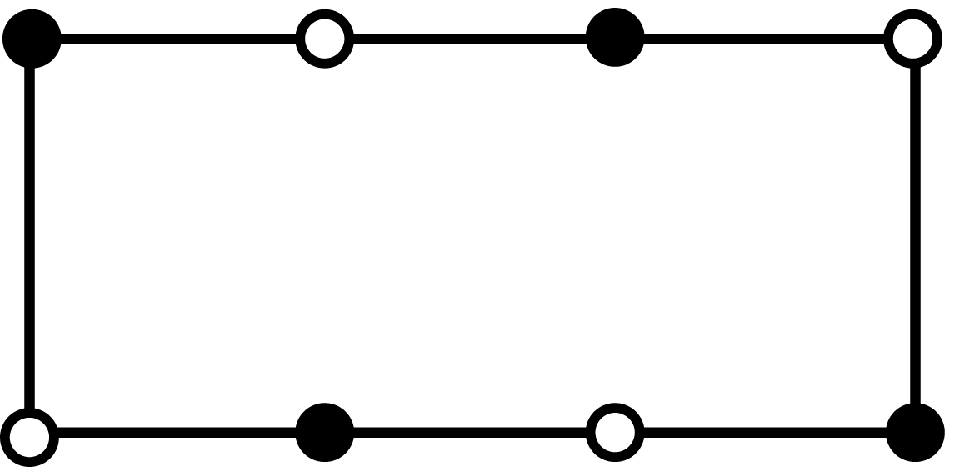}& \includegraphics[height=.9cm]{c8} &  \includegraphics[height=1.6cm]{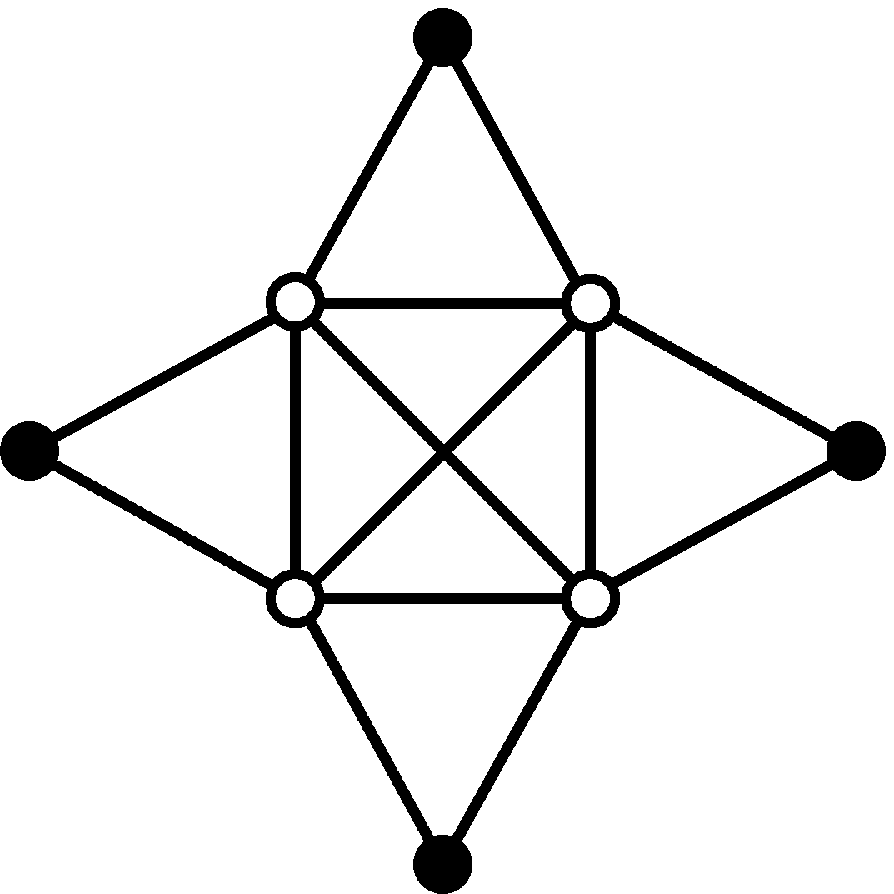}& \includegraphics[height=1.6cm]{s4}
\\

$C_8$& $C_8$ & $S_4$ &$S_4$\\
\hline
 ~ &~ &~ &~\\
 \includegraphics[height=1cm]{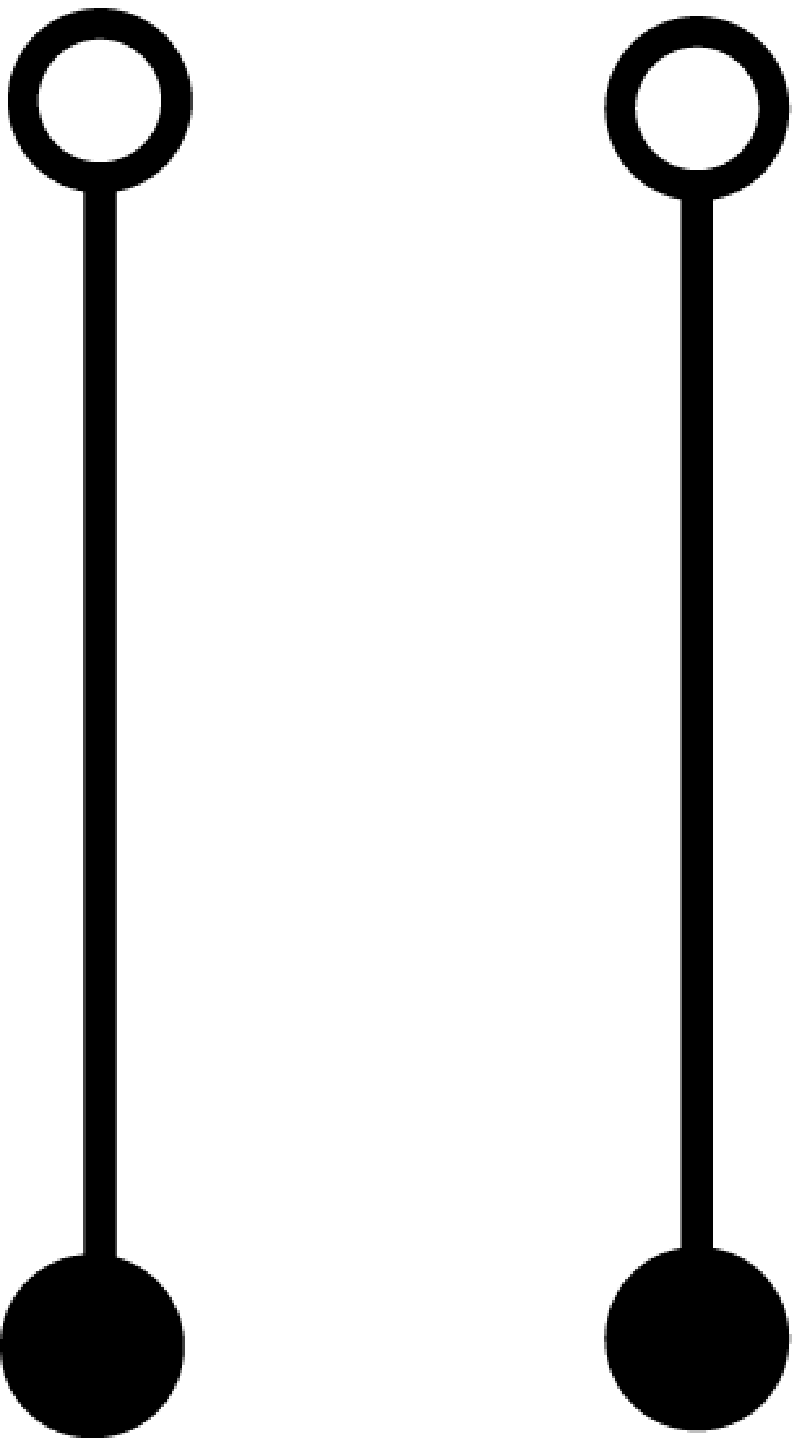}& \includegraphics[height=1cm]{2k2} &  \includegraphics[height=.9cm]{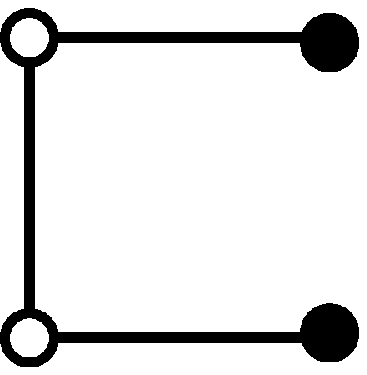}&  \includegraphics[height=.9cm]{splitp4}\\

$2K_2$& $2K_2$ & $P_4$  & $P_4$ \\
\hline
 ~ &~ &~ &~\\
\includegraphics[height=1cm]{3k2}& \includegraphics[height=1cm]{c6} &  \includegraphics[height=1.4cm]{net}&  \includegraphics[height=1cm]{s3}\\

$3K_2$ & $C_6$ & $\overline{S_3}$ & $S_3$\\
\hline

\end{tabular}
\end{center}
\caption{Correspondences between bipartite graphs, split graphs, and their mirrors.\label{fig:nets}}
\end{figure}

Figure \ref{fig:nets} illustrates the mirror of the bipartite graphs $C_6, C_8, 2 K_2,$ $ 3 K_2$ and their split graphs $S_3, S_4, P_4, \overline{S_3}$.

Note that $mir(S_3)=\overline{S_3}$ and $mir(S_4)=S_4$. Moreover, the following equalities obviously hold:

\begin{proposition}\label{prop:mirsplit}
\mbox{ }
\begin{enumerate}
\item[$(1)$] For any bipartite graph $B=(X,Y,E)$, $\overline{split_X(B)} = mir(split_Y(B))$.
\item[$(2)$] For any split graph $G=(K,I,E)$, $mir(bip(G))= bip(\overline{G})$ as well as $\overline{G}=split_I(mir(bip(G)))$.
\end{enumerate}
\end{proposition}

Recall that a bipartite graph is {\em auto-chordal bipartite} (ACB for short) if $B$ and $mir(B)$ are chordal bipartite.
Since for every $k \ge 5$, $C_{2k}$ contains an induced subgraph $3K_2$, it follows:

\begin{proposition}\label{strchcostrchforbiddensubgrchar}
A bipartite graph $B$ is an ACB graph if and only if $B$ is $(3K_2,C_6,C_8)$-free.
\end{proposition}

Strongly chordal graphs whose complement graph is strongly chordal as well are called \textit{auto-strongly-chordal} in this paper.
The next proposition follows from Lemma \ref{lemma:splitgrchar} and the following three facts: Auto-strongly-chordal graphs are exactly the $S_k$-free and $\overline{S_k}$-free split graphs; for $k \ge 5$, sun $S_k$ contains net $\overline{S_3}$ as induced subgraph; $S_4 \sim \overline{S_4}$.

\begin{proposition}
The following are equivalent:
\begin{enumerate}
\item[$(i)$] $G$ is auto-strongly-chordal.
\item[$(ii)$] $G$ is a $(S_3,\overline{S_3},S_4)$-free split graph.
\item[$(iii)$] $G$ is $(2K_2,C_4,C_5,S_3,\overline{S_3},S_4)$-free.
\end{enumerate}
\end{proposition}
Together with Lemma \ref{lemma:cbgstrchsplit} this gives:

\begin{corollary}\label{ACBautostronglych}
A bipartite graph $B=(X,Y,E)$ is an ACB graph if and only if $split_X(B)$ $(split_Y(B)$, respectively$)$ is auto-strongly-chordal.
\end{corollary}

Summarizing, we obtain:

\begin{corollary}\label{char:ACB_forbidden}
For a bipartite graph $B=(X,Y,E)$, the following are equivalent:
\begin{enumerate}
\item[$(i)$] $B$ is an ACB graph.
\item[$(ii)$] $B$ is $(3K_2,C_6,C_8)$-free.
\item[$(iii)$] $B$ is $3K_2$-free chordal bipartite.
\item[$(iv)$] $split_X(B)$ and $\overline{split_X(B)}$ are strongly chordal.
\end{enumerate}
\end{corollary}

\section{Dilworth Number of Hypergraphs, Graphs, Bipartite Graphs and Split Graphs}\label{Sect:Dilworth}

\subsection{Dilworth Number and Poset Width}

The {\em width} of a poset is the maximum number of its pairwise incomparable elements. In \cite{Spinr2003}, Chapter 8.5 deals with the Dilworth number of graphs and the width of posets. We need the following notions of the Dilworth number:

\subsubsection{Dilworth Number of Hypergraphs}

Let $H=(V,{\cal E})$ be a hypergraph. The {\em Dilworth number} $dilw(H)$ of $H$ is the maximum number of pairwise incomparable hyperedges in ${\cal E}$ with respect to set inclusion. Obviously:
$$dilw(H)=dilw(mir(H)).$$
Note that hypergraphs with pairwise incomparable hyperedges are also called {\em Sperner hypergraphs}.
A Sperner hypergraph $H=(V,{\cal E})$ is {\em $k$-critical} if $H$ has $k$ hyperedges and if for all $v \in V$, deleting $v$ in $H$ leads to a non-Sperner hypergraph.

\subsubsection{Dilworth Number of Graphs}

For graphs, the {\em vicinal preorder} $\preceq$ on the vertex set of a graph $G$ is defined as
\begin{enumerate}
\item[ ] $x \preceq y$ if $N(x) \subseteq N[y]$.
\end{enumerate}
The {\em Dilworth number} $dilw(G)$ of $G$ is the maximum integer $k$ such that there are
$k$ pairwise incomparable nodes with respect to $\preceq$ in $G$. Obviously, $dilw(G)=dilw(\overline{G})$ holds \cite{FoeHam1977/2}. In \cite{FoeHam1977/2}, for a subset $S \subseteq V$, the Dilworth number $\nabla_G(S)$ with respect to $S$ is defined as the maximum number of elements of $S$ that are pairwise incomparable in the vicinal preorder of $G$.

\subsubsection{Dilworth Number of Bipartite Graphs}

In a bipartite graph $B=(X,Y,E)$, for every $x \in X$ and $y \in Y$, the neighborhoods $N(x)$ and $N(y)$ are incomparable. Thus, it only makes sense to compare neighborhoods of vertices from $X$ (from $Y$, respectively). Moreover, for $x,x' \in X$, $N(x) \subseteq N[x']$ if and only if $N(x) \subseteq N(x')$. Thus, the vicinal preorder for bipartite graphs can be defined as follows:
\begin{enumerate}
\item[$(i)$] for $x,x' \in X$, $x \preceq x'$ if $N(x) \subseteq N(x')$ and analogously,
\item[$(ii)$] for $y,y' \in Y$, $y \preceq y'$ if $N(y) \subseteq N(y')$.
\end{enumerate}
Let $\nabla_B(X)$ ($\nabla_B(Y)$, respectively) be the Dilworth number of the corresponding neighborhood hypergraph
${\cal N}_X=\{N(x): x \in X\}$ (${\cal N}_Y=\{N(y): y \in Y\}$, respectively), which we will refer to as the \emph{$X$-Dilworth number} (\emph{$Y$-Dilworth number}, respectively).
We define the \textit{bipartite Dilworth number} of $B$ as
$$bip-dilw(B):= \max(\nabla_B(X),\nabla_B(Y)).$$

Obviously, for a bipartite graph $B=(X,Y,E)$, $bip-dilw(B) \leq \nabla_B(X) + \nabla_B(Y)$ holds.
The Dilworth number $dilw(B)$ of a bipartite graph $B$ can be as large as the sum of the $X$- and $Y$-Dilworth numbers of $B$, as is the case for the $C_{2k}$ with $k \ge 3$.

\subsubsection{Dilworth Number of Split Graphs}

For a split graph $G=(K,I,E)$, the Dilworth number can be defined in a very similar way. Since $K$ is a clique, $I$ is independent, and for every $x \in K$ and $y \in I$, $N(y) \subseteq N[x]$ holds, it is natural to define $\nabla_G(K)$ ($\nabla_G(I)$, respectively) as the maximum number of pairwise incomparable neighborhoods
of vertices in $K$ (of vertices in $I$, respectively). Then, similarly as for bipartite graphs, let
$$dilw(G):=\max(\nabla_G(K),\nabla_G(I)).$$
Thus, for a split graph $G$ and its bipartite version $bip(G)$, $dilw(G)=bip-dilw(bip(G))$ holds and is the same as the maximum of the Dilworth numbers of the corresponding neighborhood hypergraphs.

\subsection{Related Work on Small Dilworth Numbers}\label{relwork}

\subsubsection{Dilworth Number 1}\label{relworkDilw1}

Recall that a bipartite graph is called a chain graph if it is $2K_2$-free. Obviously, the following holds:

\begin{proposition}
A bipartite graph $B=(X,Y,E)$ is a chain graph if and only if $\nabla_B(X)=\nabla_B(Y)=1$.
\end{proposition}

The corresponding class for split graphs is the class of threshold graphs; in \cite{ChvHam1977}, it is shown:

\begin{proposition}\label{prop:TSgraph}
Let $G$ be a split graph. The following are equivalent:
\begin{itemize}
\item[$(i)$] $G$ is a threshold graph.
\item[$(ii)$] $dilw(G) = 1$.
\item[$(iii)$] $G$ is $(2K_2,C_4,P_4)$-free.
\end{itemize}
\end{proposition}

Note that $split_X(2K_2)=P_4$.

\subsubsection{Dilworth Number 2 for Split Graphs}\label{relworkDilw2}

Interval graphs form a famous graph class; it is well-known that these graphs are strongly chordal, and a graph is an interval graph if and only if it is chordal and its complement graph is a comparability graph (see e.g. \cite{BraLeSpi1999}). Since a graph $G$ is a permutation graph if and only if $G$ and $\overline{G}$ are comparability graphs, it follows that $G$ and $\overline{G}$ are interval graphs if and only if $G$ is a split graph and a permutation graph. Various papers deal with such graphs \cite{BenHamDeW1985,KorLozMay2013}.

Let {\em rising sun} denote the graph resulting from $S_4$ by deleting one of its simplicial vertices, and let {\em co-rising sun} denote its complement graph (see Figure \ref{fig:SPLITXP7}).
F\"oldes and Hammer \cite{FoeHam1977/2} proved the following property:

\begin{proposition}[\cite{FoeHam1977/2}]\label{splitintervchar}
A split graph $G$ is an interval graph if and only if $G$ is $(S_3, \overline{S_3},rising~sun)$-free.
\end{proposition}

\begin{corollary}[\cite{BenHamDeW1985}]\label{intervcointervchar}
If $G$ is a split graph then $G$ and $\overline{G}$ are both interval graphs if and only if $G$ is $(S_3,\overline{S_3},rising~sun,co$-$rising~sun)$-free.
\end{corollary}

As already mentioned, for split graphs $G=(K,I,E)$, F\"oldes and Hammer in \cite{FoeHam1977/2} defined the $I$-Dilworth number $\nabla_G(I)$ for the independent set $I$. They showed:
\begin{proposition}[\cite{FoeHam1977/2}]
\label{splitDilw}
Let $G=(K,I,E)$ be a split graph. Then $G$
is an interval graph if and only if $\nabla_G(I) \le 2$.
\end{proposition}

\begin{corollary}
A graph $G$ and its complement are interval graphs if and only if the Dilworth number of $G$ is at most $2$.
\end{corollary}

Thus, $S_3$, $\overline{S_3}$, rising sun, and co-rising sun are the minimal split graphs of Dilworth number~3.

\medskip

\noindent
{\bf Remark.} Surprisingly, both papers \cite{FoeHam1977/2} and \cite{BenHamDeW1985} have nearly the same title, namely ``Split graphs of Dilworth number 2'', and in Theorem 5 of \cite{BenHamDeW1985}, the result of \cite{FoeHam1977/2} is cited in a wrong way, namely as ``$G$ is an interval and split graph if and only if its Dilworth number is at most 2'' (but the results in \cite{BenHamDeW1985} are correct).

\subsubsection{Dilworth Number 2 for Bipartite Graphs}\label{relworkDilw2bip}

\begin{figure}
\begin{center}
\begin{tabular}{cccc}
 \includegraphics[width=1.8cm]{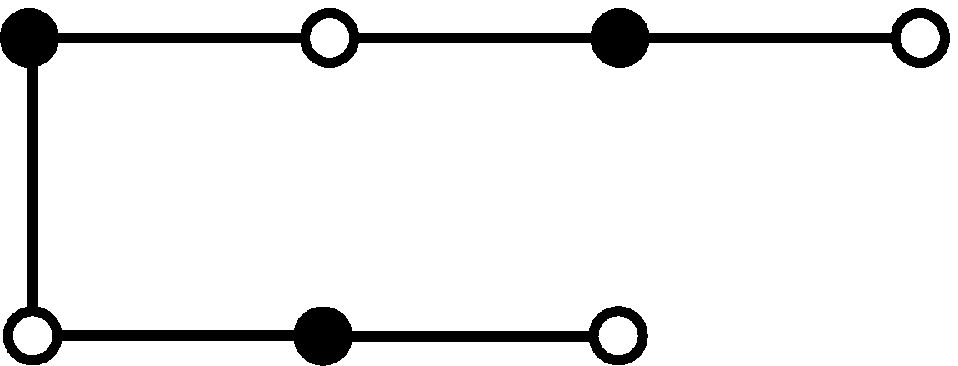}& \includegraphics[width=1.8cm]{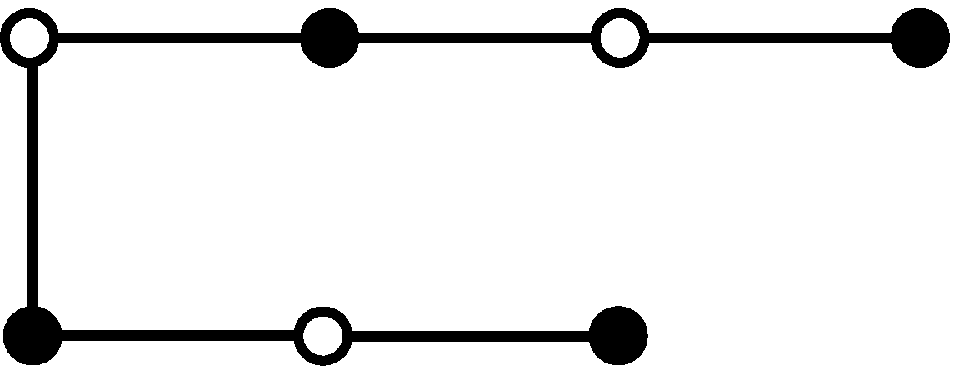} &  \includegraphics[height=1.3cm]{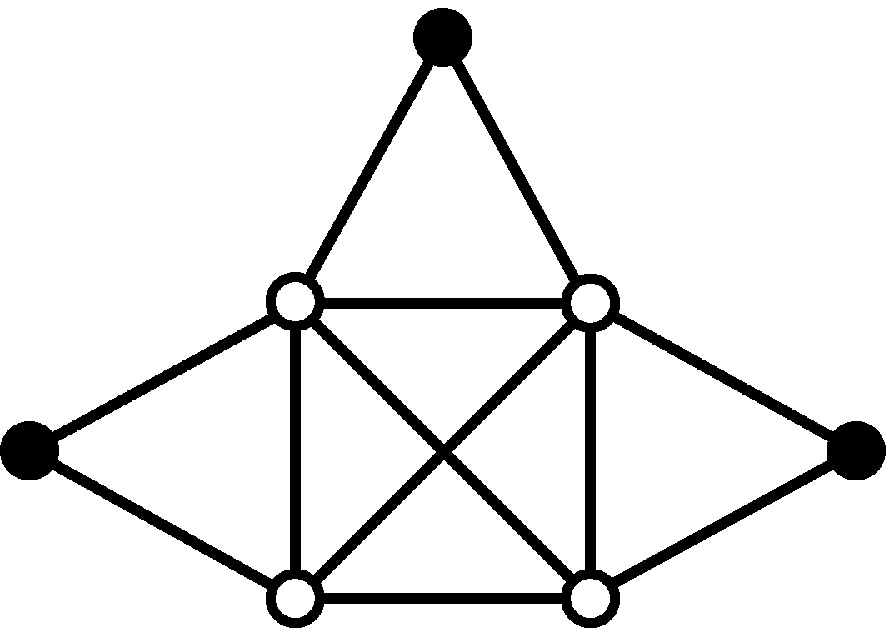}&  \includegraphics[height=.7cm]{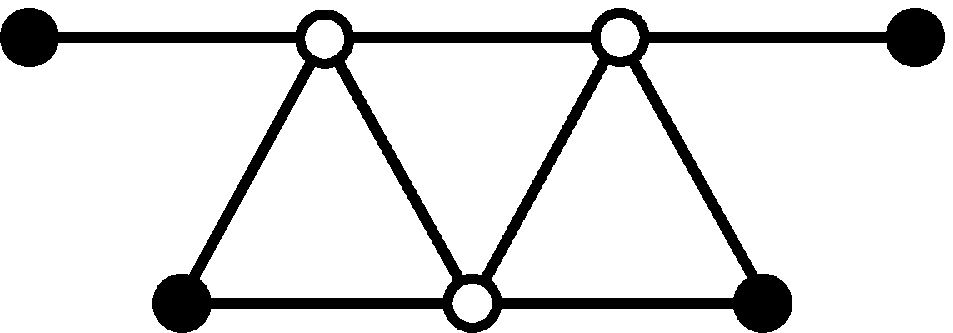}\\
$X$-$P_7$ & $Y$-$P_7$ &$split_X(X$-$P_7)$&   $split_X(Y$-$P_7)$\\
&&rising sun&co-rising sun\\
\end{tabular}
\end{center}
\caption{Correspondences between the $P_7$s and their split graphs.\label{fig:SPLITXP7}}

\end{figure}

Recall that $split_X(C_6)=S_3$ and $split_X(3K_2)=\overline{S_3}$. Obviously, for an $X$-$P_7$ $P$, $split_X(P)$ is the rising sun, and for a $Y$-$P_7$ $P$, $split_X(P)$ is the co-rising sun,
as illustrated by Figure \ref{fig:SPLITXP7}.

Split graphs which are interval graphs are characterized in Proposition \ref{splitintervchar} as being $(S_3, \overline{S_3}$, $rising$ $sun)$-free and in Proposition \ref{splitDilw} as having $I$-Dilworth number at most 2.
Let us translate Proposition \ref{splitintervchar} into terms of ACB graphs.

\begin{proposition}\label{splitinterv3K2C6XP7}
For a bipartite graph $B=(X,Y,E)$, $split_X(B)$ is an interval graph if and only if $B$ is $(3K_2,C_6,X$-$P_7)$-free.
\end{proposition}

Since $C_8$ contains $P_7$, it follows:

\begin{corollary}
For an ACB graph $B=(X,Y,E)$, $split_X(B)$ is an interval graph if and only if $B$ is $X$-$P_7$-free.
\end{corollary}
Corollary \ref{intervcointervchar} corresponds to a more restricted class of ACB graphs:

\begin{proposition}\label{intervcointerv3K2C6P7}
Let $B=(X,Y,E)$ be a bipartite graph. The following are equivalent:
\begin{itemize}
\item[$(i)$] $B$ is $(3K_2,C_6,P_7)$-free.
\item[$(ii)$] $B$ is a $P_7$-free ACB graph.
\item[$(iii)$] $split_X(B)$ and $split_Y(B)$ are both interval graphs.
\end{itemize}
\end{proposition}

Recall that for a split graph $G=(K,I,E)$ with clique $K$ and stable set $I$, $bip(G)$ denotes the bipartite graph resulting from $G$ by turning $K$ into a stable set.
\begin{proposition}
Let $G$ be a split graph. The following are equivalent:
\begin{itemize}
\item[$(i)$] $G$ is $(S_3$,$\overline{S_3}$,rising sun,co-rising sun$)$-free.
\item[$(ii)$] $bip(G)$ is $(3K_2,C_6,P_7)$-free.
\item[$(iii)$] $bip(G)$ is a $P_7$-free ACB graph.
\end{itemize}
\end{proposition}

In \cite{BenHamDeW1985}, a linear-time recognition algorithm for split graphs of Dilworth number greater than 2 is given.
This approach can easily be adapted to decide whether a bipartite graph has $X$-Dilworth number ($Y$-Dilworth number, respectively) more than 2.

\subsubsection{Dilworth Numbers 3 and 4 for Split Graphs}\label{relworkDilw3}

In \cite{Nara1982}, Nara characterized split graphs of Dilworth number at most 3 by a list of forbidden induced subgraphs $G_1,\ldots,G_{16}$ which, together with their complement graphs, represent all $4$-critical split graphs; for $k \ge 2$, a split graph $G=(K,I,E)$ is called {\em $k$-critical} (with respect to $I$) in \cite{Nara1982} if $\nabla_G(I) = k$ and for all $v \in I$, $\nabla_{G-v}(I-v) \le k-1$.

\begin{theorem}[\cite{Nara1982}]\label{splitgrDilw3}
The Dilworth number of a split graph $G$ is at most $3$ if and only if $G$ and $\overline{G}$ are $(G_1,\ldots,G_{16})$-free.
\end{theorem}

It should be noted that the graph $G_{11}$ of Nara's list in \cite{Nara1982} contains the 4-sun $S_4$ as an induced subgraph (which is graph $G_{12}$ of his list) and thus is not 4-critical. But we verified Theorem \ref{splitgrDilw3} by determining all 4-critical (and 5-critical) Sperner hypergraphs using a well adapted backtracking. It turned out that there are 15 4-critical and 178 5-critical Sperner hypergraphs.

\begin{figure}
\begin{center}
\begin{tabular}{ccc}
\includegraphics[height=1.3cm]{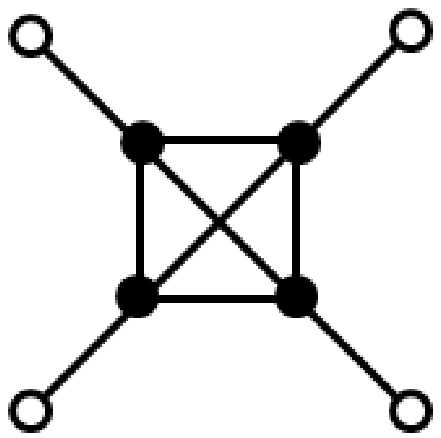}&\includegraphics[height=1.3cm]{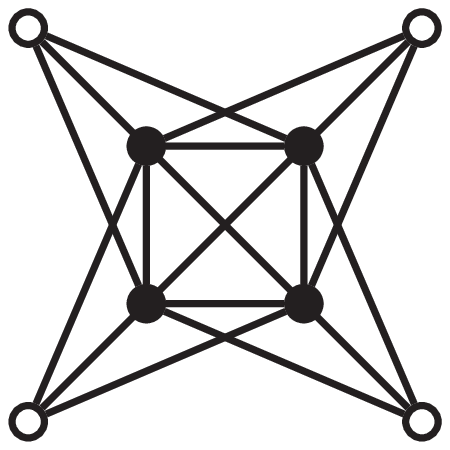}&\includegraphics[height=1.3cm]{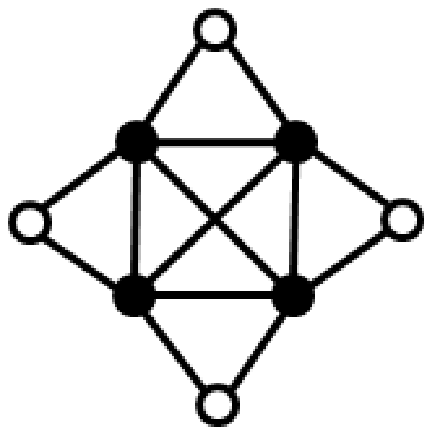}\\
$H_1$&mirror of $H_1$&$H_2$ (auto-mirror)\\
\includegraphics[height=1.3cm]{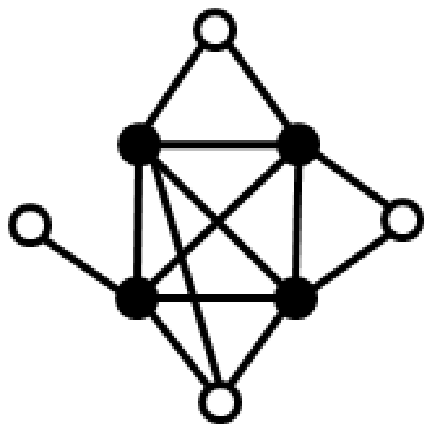}&\includegraphics[height=1.3cm]{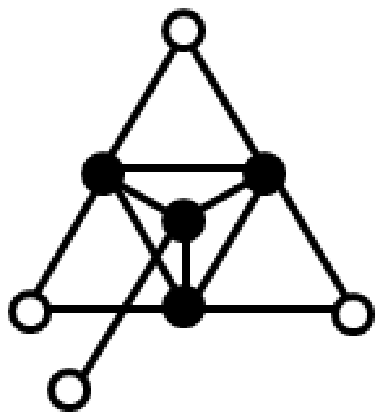}&\includegraphics[height=1.3cm]{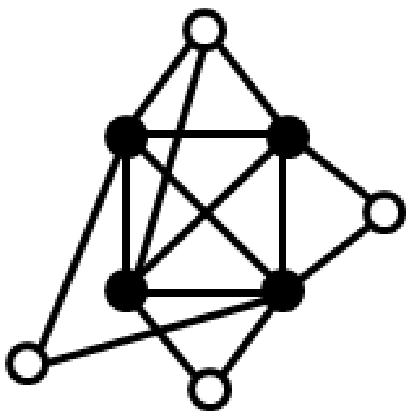}\\
$H_3$ (auto-mirror)&$H_4$&mirror of $H_4$\\
\includegraphics[height=1.3cm]{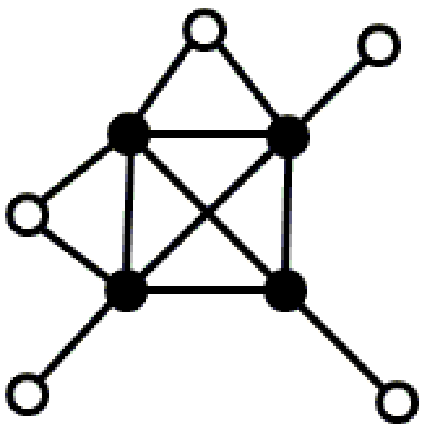}&\includegraphics[height=1.3cm]{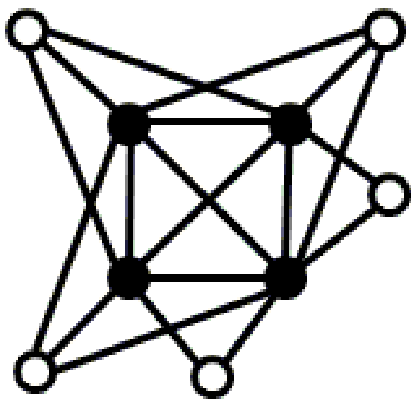}&~\\
$H_5$& mirror of $H_5$&~\\
\includegraphics[height=1.3cm]{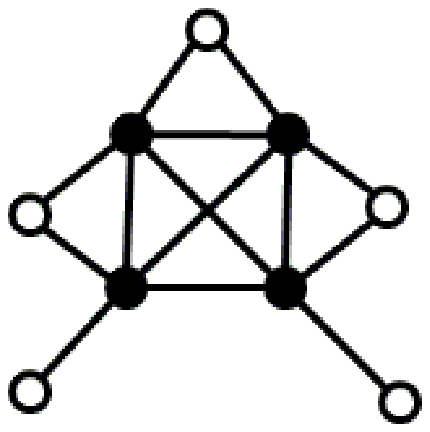}&\includegraphics[height=1.3cm]{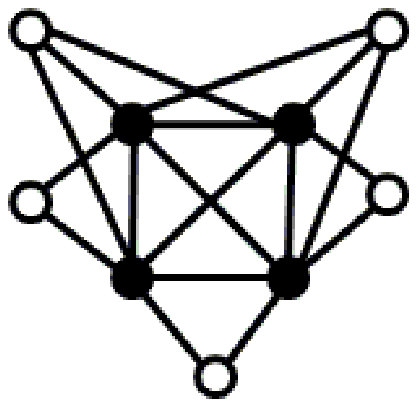}&~\\
$H_6$& mirror of $H_6$&~\\
\includegraphics[height=1.3cm]{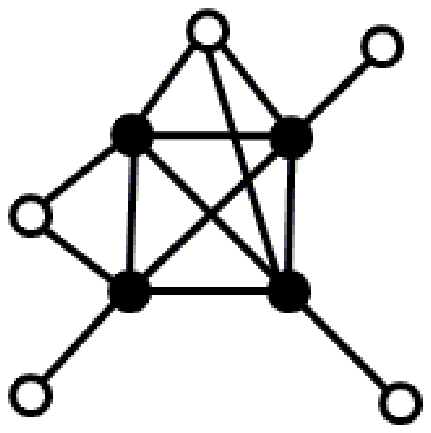}&\includegraphics[height=1.3cm]{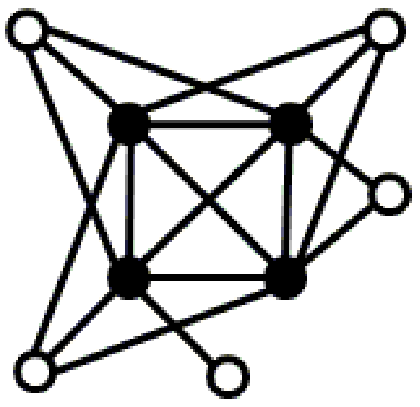}&~\\
$H_7$& mirror of $H_7$&~\\
\includegraphics[height=1.3cm]{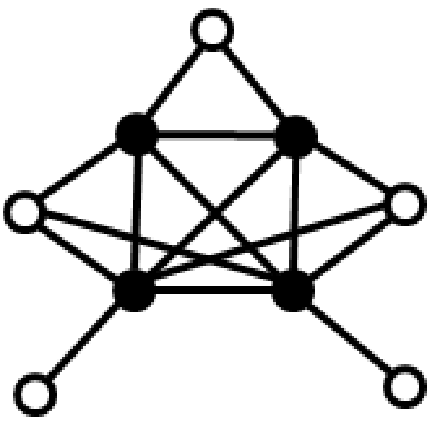}&\includegraphics[height=1.3cm]{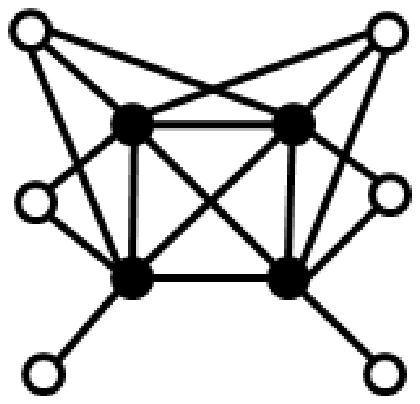}&\includegraphics[height=1.3cm]{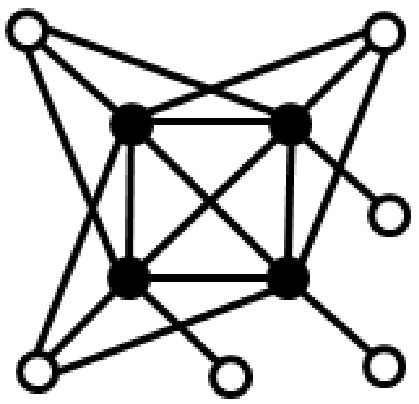}\\
 $H_8$ (auto-mirror)&$H_9$ (auto-mirror) & $H_{10}$ (auto-mirror)
\end{tabular}
\end{center}
\caption{4-critical split graphs $H_1-H_{10}$ and their mirrors.\label{fig:H1-H10}}
\end{figure}

Up to $G_{11}$ all other 15 graphs of Nara's list are determined by the 4-critical Sperner hypergraphs and hence are 4-critical.
The 4-critical split graphs $H_1-H_{10}$ and their mirrors in Figure \ref{fig:H1-H10} represent the 15 4-critical Sperner hypergraphs.

\medskip

Theorem \ref{splitgrDilw3} can also be formulated in the following way:
\begin{theorem}\label{splitgrDilw3H1-H10}
The Dilworth number of a split graph $G$ is at most $3$ if and only if $G$ and $\overline{G}$ are $(H_1,\ldots,H_{10})$-free and $(mir(H_1),\ldots,mir(H_{10}))$-free .
\end{theorem}

\section{Arbitrarily large $X$- and $Y$-Dilworth Numbers of ACB Graphs}\label{Sect:DilwACB-XY}

In the following, we show that for ACB graphs $G$, both $\nabla_G(X)$ and $\nabla_G(Y)$ can be arbitrarily large.
For $k \ge 2$, let $D_k=(X,Y,E)$ be the bipartite graph with vertex sets $X=X_{3k-4}= \{x_1,\dots,x_{3k-4}\}$ and $Y=Y_{3k-4}=\{y_1,\dots,y_{3k-4}\}$ where $x_iy_j$ is an edge if and only if
\[
|i-j|\le k-2 \text{ and }\{i,j\} \cap \{k-1,\dots,2k-2\}\neq  \emptyset.
\]

\begin{figure}
 \begin{center}
\includegraphics[height=3cm]{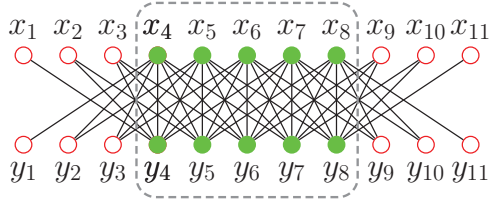}
 \end{center}
\caption{Graph $D_5$ having $X$- and $Y$-Dilworth number 5: $\nabla_B(X) = \nabla_B(Y) =5$.\label{fig:D5}}
\end{figure}

Figure \ref{fig:D5} shows graph $D_5$.
In order to show the subsequent Theorem \ref{ACBDilwk}, we color the vertices $x_i$ and $y_j$ {\em green} if $i,j \in \{k-1,\dots,2k-2\}$ and {\em red} otherwise.  Note that in $D_k$, $X$ consists of a green central interval of $k$ vertices surrounded by two red intervals, each of $k-2$ vertices, and the same holds for $Y$. Note also that each edge of $D_k$ has at least one green vertex, and subgraphs induced by one of the red intervals and its green neighbors form chain graphs, i.e., are $2K_2$-free.

\begin{theorem}
\label{ACBDilwk}
For every $k \ge 2$, $D_k$ is an ACB graph with Dilworth numbers $\nabla_{D_k}(X)=\nabla_{D_k}(Y)=k$.
\end{theorem}

\noindent
{\em Proof.} Obviously, $|N(x_i)|=2(k-2)+1$ if $i \in \{k-1,\dots,2k-2\}$. Moreover,
$N(x_1) \subseteq \dots \subseteq N(x_{k-1})$ and $N(x_{2k-2}) \supseteq \dots \supseteq N(x_{3k-4})$.
Thus $\nabla_{D_k}(X)=k$ and by symmetry also $\nabla_{D_k}(Y)=k$.

In order to show that $D_k$ is an ACB graph, we need some lemmas:
\begin{lemma}
\label{lem1}
Let $x_{i_1}y_{j_1}, x_{i_2}y_{j_2}$ be edges but $x_{i_1}y_{j_2}$ and $x_{i_2}y_{j_1}$ not be edges in $D_k$. Then  $i_1 < i_2$ implies  $j_1 < j_2$.
\end{lemma}

\noindent
{\em Proof.} Assume to the contrary that
\begin{equation}
\label{ij}
i_1 < i_2 \text{ and } j_1 > j_2.
\end{equation}

By symmetry, we may assume that $x_{i_1}$ is green. Since $x_{i_1}y_{j_1}, x_{i_2}y_{j_2}$ are edges
\begin{align*}
i_1-(k-2) &\le j_1 \le i_1+(k-2),\\
i_2-(k-2) &\le j_2 \le i_2+(k-2)
\end{align*}
and consequently in view of (\ref{ij})
\[
i_1-(k-2) < i_2-(k-2) \le j_2 < j_1 \le i_1+(k-2),
\]
i.e., $x_{i_1}y_{j_2}$ is an edge, a contradiction.
\qed

\begin{lemma}
\label{lem2}
Let $x_{i}y_{j_1}, x_{i}y_{j_3}$ be edges in $D_k$. Then $j_1 < j_2 < j_3$ implies that $x_iy_{j_2}$ is an edge in $D_k$.
\end{lemma}

\noindent
{\em Proof}. Let $j_1 < j_2 < j_3$.

\medskip

\noindent
{\bf Case 1}. $x_i$ is green. By supposition, the interval $\{x: |i-x| \le k-2\}$ contains $j_1,j_3$ and hence it contains also $j_2$. But this implies that $x_iy_{j_2}$ is an edge.

\medskip

\noindent
{\bf Case 2}. $x_i$ is red. Then $y_{j_1}$ and $y_{j_3}$ are green and hence also $y_{j_2}$ is green.
By symmetry, we may assume that $i < k-1$.
By supposition, the interval $\{x: k-1 \le x \le i+(k-2)\}$ contains $j_1,j_3$ and hence it contains also $j_2$. But this implies that $x_iy_{j_2}$ is an edge.
\qed

\begin{lemma}
The graph $D_k$ is $3 K_2$-free.
\end{lemma}

\noindent
{\em Proof}. Assume the contrary and let $x_{i_1}y_{j_1}, x_{i_2}y_{j_2}, x_{i_3}y_{j_3}$ be the edges of an induced $3 K_2$. We may assume that $i_1 < i_2 < i_3$. Then, by Lemma \ref{lem1}, $j_1 < j_2 < j_3$.
First we show that $x_{i_2}$ is green. Indeed, assume that it is red. By symmetry we may assume that $i_2 < k-1$. But then also $i_1 < k-1$ which means that $x_{i_1}$ is red. Then $y_{j_1}$ and $y_{j_2}$ have to be green. Since $x_{i_1}y_{j_1}, x_{i_2}y_{j_2}$ are edges,
\begin{align*}
k-1 &\le j_1 \le i_1+(k-2),\\
k-1 &\le j_2 \le i_2+(k-2).
\end{align*}
With $i_1 < i_2$ it follows that
\[
k-1 \le j_1 \le i_1+(k-2) < i_2 + (k-2),
\]
i.e., $x_{i_2}y_{j_1}$ is an edge, a contradiction.

\medskip

Thus $x_{i_2}$ and analogously $y_{j_2}$ are green. By symmetry we may assume that $x_{i_1}$ is green.
Then $k-1 \le i_1 < i_2 \le 2k-2$ and $k-1 \le j_2 \le 2k-2$.

\medskip

\noindent
{\bf Case 1.} $j_2 < 2k-2$. Then
\[
-(k-2) = (k-1)-(2k-3)\le j_2-i_1 \le (2k-3)-(k-1)=k-2,
\]
i.e., $x_{i_1}y_{j_2}$ is an edge, a contradiction.

\medskip

\noindent
{\bf Case 2.} $j_2=2k-2$. Then $y_{j_3}$ is red because of $j_2 =2k-2 < j_3$. It follows that $x_{i_3}$ is green, i.e., $k-1 \le i_1 < i_2 < i_3 \le 2k-2$, and further
\[
-(k-2) \le (2k-2)-(2k-2) \le j_2-i_3 \le (2k-2)-(k+1)< k-2,
\]
i.e., $x_{i_3}y_{j_2}$ is an edge, a contradiction.
\qed

\medskip

Now the proof of Theorem \ref{ACBDilwk} is completed by the following:

\begin{lemma}
The graph $D_k$ is $C_6$-free and $C_8$-free.
\end{lemma}

\noindent
{\em Proof}. Assume that $D_k$ contains an induced $C_6=x_{i_1}y_{j_1}x_{i_2}y_{j_2}x_{i_3}y_{j_3}x_{i_1}$.
Let $\{l_1,l_2,l_3\}=\{j_1,j_2,j_3\}$ where $l_1 < l_2 < l_3$, i.e., the ``$l$-set'' is the ``ordered $j$-set''. Since the degrees of the vertices $x_{i_1}, x_{i_2}, x_{i_3}$ in the cycle $C_6$  are exactly 2, by Lemma \ref{lem2} the neighborhoods of $x_{i_1}, x_{i_2}, x_{i_3}$ have to be either $\{y_{l_1},y_{l_2}\}$ or $\{y_{l_2},y_{l_3}\}$, hence two of these vertices have the same neighborhood in the cycle, a contradiction. In the same way one can show that $D_k$ does not contain a $C_8$.
\qed

\medskip

Theorem \ref{ACBDilwk} states that ACB graphs and hence also auto-strongly-chordal graphs have unbounded Dilworth number.

\begin{corollary}\label{ACBDilworthkl}
For all $k,l \ge 2$, there is an ACB graph $G=(X,Y,E)$ with Dilworth numbers $\nabla_{G}(X)=k$ and $\nabla_{G}(Y)=l$.
\end{corollary}
\proof
In the case $k=l$, the graph $D_k$ fulfills Corollary \ref{ACBDilworthkl} by Theorem \ref{ACBDilwk}.
Thus we may assume that $k > l \ge 2$.
Let $D_{k,l}$ be the graph resulting from $D_k$ by omitting the $k-l$ green vertices $y_k,\dots,y_{2k-l-1}$ (thus, keeping only $l$ green vertices in $Y$). As an induced subgraph of an ACB graph it is an ACB graph as well.
As for $D_k$, we have also in this graph $N(x_1) \subseteq \dots  \subseteq N(x_{k-1})$,
$N(x_{2k-2}) \supseteq \dots  \supseteq N(x_{3k-4})$, $N(y_1) \subseteq \dots  \subseteq N(y_{k-1})$,
$N(y_{2k-2}) \supseteq \dots  \supseteq N(y_{3k-4})$. Moreover, the neighborhoods of
$x_{k-1},\dots,x_{2k-2}$ are distinct and of the same size. Hence $\nabla_{D_{k,l}}(X)=k$.
Similarly, the neighborhoods of the $l$ vertices
$y_{k-1},y_{2k-l},y_{2k-l+1},\dots,y_{2k-2}$ are distinct and of the same size. Hence $\nabla_{D_{k,l}}(Y)=l$. \qed
\medskip

\begin{figure}
\begin{center}
 \includegraphics[height=5.1cm]{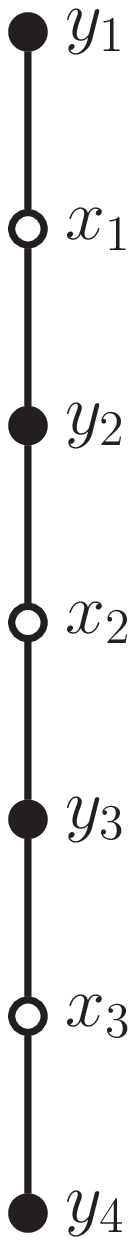} ~~~~~~~~~~~~~~~~~~
   \includegraphics[height=4.5cm]{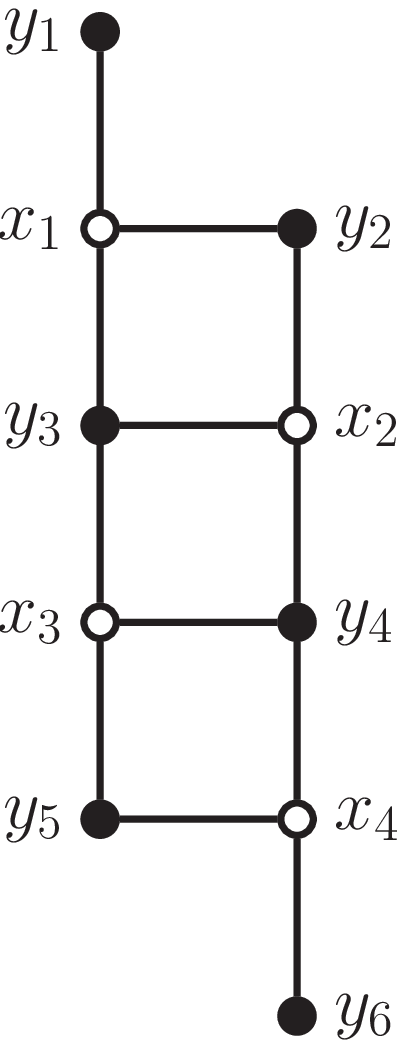}
 \end{center}
\caption{Critical bipartite graphs $B_3$ and $B_4$ with $\nabla_{B_3}(X)=3$, $\nabla_{B_3}(Y)=2$, $\nabla_{B_4}(X)=4$, and $\nabla_{B_4}(X)=2$.\label{fig:B3B4}}
\end{figure}

In the next section, the following graphs play an essential role: Let $B_k$ denote the bipartite graph with vertex sets $X_k=\{x_1,\ldots,x_k\}$ and $Y_k=\{y_1,\ldots,y_{2k-2}\}$ such that for every $i \in \{1,\ldots,k\}$, $N(x_i)=\{y_i,\ldots,y_{i+k-2}\}$.
Note that $B_2$ is the $2K_2$, $B_3$ is the $P_7$, which is the bipartite counterpart of the rising sun ($X$-$P_7$) and of the co-rising sun ($Y$-$P_7$), and
$B_4$ is the bipartite counterpart of graph $G_5$ from \cite{Nara1982}. Figure \ref{fig:B3B4} illustrates the bipartite graphs $B_3$ and $B_4$.

\begin{corollary}
\label{DilwBk}
For every $k \ge 2$, $B_k$ is an ACB graph with $\nabla_{B_k}(X)=k$ and $\nabla_{B_k}(Y)=2$.
\end{corollary}

\proof Obviously, for all $k \ge 2$, $B_k$ is an induced subgraph of $D_k$, namely induced by the vertices $x_{k-1},\ldots, x_{2k-2}$ and $y_1,\ldots,y_{k-1},y_{2k-2},\ldots,y_{3k-4}$. By Theorem \ref{ACBDilwk}, this implies that $B_k$ is an ACB graph. Moreover, clearly, $\nabla_{B_k}(X_k) \le \nabla_{D_k}(X_k)=k$. In $B_k$, for every $i \neq j$, the neighborhoods of $N(x_i)$ and $N(x_j)$ are incomparable because they are distinct and of the same size. Thus $\nabla_{B_k}(X)=k$, and obviously $\nabla_{B_k}(Y)=2$.
\qed

\section{Characterization of ACB Graphs with bounded Dilworth Number}\label{Sect:DilwACBbounded}

Let $k \ge 2$. An ACB graph $G=(X,Y,E)$ is called \emph{$k$-critical} (with respect to Dilworth number) if $bip-dilw(G)=k$ and for all $v \in X \cup Y$, $bip-dilw(G-v)\le k-1$
holds. The main result of this section is Theorem \ref{ACBDilwkchar} which shows that the $k$-critical ACB graphs are the graphs $B_k$, introduced in the preceding section.

The graphs $B_k$ have the following nice property:
\begin{proposition}\label{Bk-1subBk}
For every $k \ge 2$, $B_{k-1}$ is an induced subgraph of $B_k$.
\end{proposition}

\noindent
{\bf Proof.}
For all $i \in \{1,\ldots,k-1\}$, $N(x_i)$ contains $y_{k-1}$. Thus, if one deletes $y_{k-1},y_{2k-2}$ and $x_k$ in $B_k$, the resulting subgraph is isomorphic to $B_{k-1}$.
\qed

\medskip

\noindent
Now we are ready to state our main theorem:
\begin{theorem}\label{ACBDilwkchar}
The bipartite graph $G$ is a $k$-critical ACB graph if and only if $G$ is isomorphic to $B_k$.
\end{theorem}

\noindent
{\bf Proof.}
By Corollary \ref{DilwBk}, $B_k$ is an ACB graph. It is
easy to check that $B_k$ is $k$-critical. This proves the if-part. In order to prove the only-if-part let $G=(X,Y,E)$ be a $k$-critical ACB graph. Let $\F=\{N(x): x \in X\}$ and let $\F^*=\{N(y): y \in Y\}$.
Note that $H^*=(X,\F^*)$ is the dual hypergraph of the hypergraph $H=(Y,\F)$.

Let without loss of generality, $dilw(Y,\F)=k$. Since $G$ is $k$-critical, $|X|=k$. Moreover, $\F$ as well as $\F^*$ do not contain multiple members, $G$ does not contain isolated vertices and the members of $\F$ are pairwise incomparable with respect to inclusion.

\begin{lemma}\label{lm1}
We have $dilw(X,\F^*)=2$.
\end{lemma}

\noindent
{\em Proof.} First assume that $dilw(X,\F^*)=1$. Let $\F^*=\{X_1,\dots,X_l\}$ with $X_i=N(y_i)$, $i=1,\dots,l$, and $X_1 \subset \dots \subset X_l$. Note that $X_l=X$. For $x \in X$, let $\iota(x)=\min\{i: x \in X_i\}$.
Let $a,b \in X$ and let, without loss of generality, $\iota(a)\le \iota(b)$. Then $\{y_{\iota(b)},y_{\iota(b)+1},\dots,y_l\}=N(b) \subseteq
N(a)=\{y_{\iota(a)},y_{\iota(a)+1},\dots,y_l\}$, a contradiction.

Now assume that  $dilw(X,\F^*)\ge 3$. Then there are $\alpha, \beta, \gamma \in Y$ such that $A=N(\alpha)$, $B=N(\beta)$ and $C=N(\gamma)$ are pairwise incomparable. We study several cases which are defined depending on the following conditions:

\begin{align}
\label{rel1}
A & \subseteq B \cup C,\\
\label{rel2}
B & \subseteq A \cup C,\\
\label{rel3}
C & \subseteq A \cup B.
\end{align}
\noindent
{\bf Case 1.} At least two of the relations (\ref{rel1}) -- (\ref{rel3}) are satisfied. Let, without loss of generality, (\ref{rel1}) and (\ref{rel2}) be true. Then $A \setminus B \subseteq C$ and $B \setminus A \subseteq C$. There is some $d \in A \cap B \setminus C$ because otherwise $A \subseteq C$ (and $B \subseteq C$). Since $A$ and $B$ are incomparable there exist elements $a \in A \setminus B$ and $b \in B \setminus A$. Then $a, \alpha, d, \beta, b, \gamma$ form a $C_6$, a contradiction.

\medskip

\noindent
{\bf Case 2.} Exactly one of the relations (\ref{rel1}) -- (\ref{rel3}) is satisfied. Let, without loss of generality, (\ref{rel1}) be true. Then $A \setminus B \subseteq C$ and there are elements $b \in B \setminus (A \cup C)$ and $c \in C \setminus (A \cup B)$. Let $a \in A \setminus B$. Since $N(a)$ and $N(c)$ are incomparable there is some $\delta \in N(c)\setminus N(a)$. We have $\delta \neq \alpha$ since $\alpha \in N(a)$ and $\delta \neq \beta$ since $\beta \notin N(c)$. Moreover, $\delta \neq \gamma$ since $\gamma \in N(a)$ in view of $a \in A \setminus B \subseteq C$.

\medskip

\noindent
{\bf Case 2.1.} $\delta \notin N(b)$. Then $a\alpha$, $b\beta$, $c\delta$ form a $3K_2$, a contradiction.

\medskip

\noindent
{\bf Case 2.2.} $\delta \in N(b)$. There is some $d \in A \cap B \setminus C$ because otherwise $A \subseteq C$.

\medskip

\noindent
{\bf Case 2.2.1} $d \notin N(\delta)$. Then $b,\beta,d,\alpha,a,\gamma,c,\delta$ form a $C_8$, a contradiction.

\medskip

\noindent
{\bf Case 2.2.2} $d \in N(\delta)$. Then $a, \gamma, c, \delta, d, \alpha$ form a $C_6$, a contradiction.

\medskip

\noindent
{\bf Case 3.} None of the relations (\ref{rel1}) -- (\ref{rel3}) is satisfied. Then there are elements $a \in A \setminus (B \cup C)$, $b \in B \setminus (A \cup C)$ and $c \in C \setminus (A \cup B)$. But $a\alpha$, $b\beta$, $c\gamma$ form a $3K_2$, a contradiction which finally shows Lemma \ref{lm1}.
\qed

\medskip

\begin{lemma}
\label{lm2}
The family $\F^*$ does not have any single maximal element.
\end{lemma}
\noindent
{\em Proof.} Assume that there is some $\alpha \in Y$ such that $N(y) \subseteq N(\alpha)$ for all $y \in Y$. Then $\nabla_{G - \alpha}(X)=k$, a contradiction.
\qed

\medskip

From Lemmas \ref{lm1} and \ref{lm2} it follows that $\F^*$ has exactly two maximal elements, say $A=N(\alpha)$ and $B=N(\beta)$.

\begin{lemma}
\label{lm3}
There is no $\gamma \in Y$ such that $N(\gamma) \subseteq A \cap B$.
\end{lemma}

\noindent
{\em Proof.} Assume that there is some $\gamma \in Y$ such that $C=N(\gamma) \subseteq A \cap B$.
Let $a \in A \setminus B$, $b \in B \setminus A$ and $c \in C$. Note that $c \neq a$ and $c \neq b$. Let $\delta \in N(a)\setminus N(c)$ and $\varepsilon \in N(b)\setminus N(c)$.

\noindent
{\bf Case 1.} $\varepsilon \notin N(a)$ and $\delta \notin N(b)$. Then $a\delta$, $c\gamma$, $b\varepsilon$ form a $3K_2$, a contradiction.

\noindent
{\bf Case 2.} $\varepsilon \in N(a)$ or $\delta \in N(b)$. Without loss of generality, we may assume that $\varepsilon \in N(a)$. Then $a, \alpha, c, \beta, b, \varepsilon$ form a $C_6$, a contra\-diction.
\qed

\medskip

Let $\F_A^*=\{C \in \F^*: C \subseteq A\}$ and $\F_B^*=\{C \in \F^*: C \subseteq B\}$. Since $A$ and $B$ are the maximal elements of $\F^*$ and in view of Lemma \ref{lm3} the families $\F_A^*$ and $\F_B^*$ form a partition of $\F^*$. \qed

\begin{lemma}
\label{lm4}
The families $\F_A^*$ and $\F_B^*$ are chains with respect to inclusion.
\end{lemma}
\noindent
{\em Proof.}  Assume e.g. that $\F_A^*$ contains two incomparable elements $C$ and $D$. Since $C \nsubseteq B$ and $D \nsubseteq B$ the members $C, D$ and $B$ are three pairwise incomparable elements of $\F^*$, a contradiction to Lemma \ref{lm1}.
\qed

\medskip

Let
\begin{align*}
\F_A^*&= \{A_1,\dots,A_r\} \text{ where } A_1 \subset \dots \subset A_r=A,\\
\F_B^*&= \{B_1,\dots,B_s\} \text{ where } B_1 \subset \dots \subset B_s=B
\end{align*}
and let $\alpha_i, \beta_j$ be those elements of $Y$ for which $N(\alpha_i)=A_i$ and $N(\beta_j)=B_j$, $i=1,\dots,r$, $j=1,\dots,s.$ Note that $Y=\{\alpha_1,\dots,\alpha_r,\beta_1,\dots,\beta_s\}$. For an element $x \in X$, let
\[
\iota_A(x)=
\begin{cases}
\min\{i: x \in A_i\}&\text{if } x \in A,\\
\infty&\text{otherwise}
\end{cases}
\]
and
\[
\iota_B(x)=
\begin{cases}
\min\{j: x \in B_j\}&\text{if } x \in B,\\
\infty&\text{otherwise.}
\end{cases}
\]

For brevity, we set $A_0=B_0=\emptyset$ and $A_{r+1}=B_{s+1}=X$.

\begin{lemma}
\label{lm5}
We have $|A_i \setminus A_{i-1}|=|B_j \setminus B_{j-1}|=1$, $i=1,\dots,r+1$, $j=1,\dots,s+1$.
\end{lemma}

\noindent
{\em Proof.}  Assume e.g. that $|A_i \setminus A_{i-1}| \ge 2$ and let $a,a'$ be two different elements of $|A_i \setminus A_{i-1}|$. Let, without loss of generality, $\iota_B(a) \le \iota_B(a')$.

\noindent
{\bf Case 1.} $\iota_B(a) = \iota_B(a') = \infty$, i.e., $a,a' \notin B$. Let, without loss of generality, $\iota_A(a)\le \iota_A(a')$. Obviously, $N(a)=\{\alpha_{\iota_A(a)}, \alpha_{\iota_A(a)+1},\dots, \alpha_r\} \supseteq \{\alpha_{\iota_A(a')},$ $ \alpha_{\iota_A(a')+1},\dots, \alpha_r\} =N(a)$, a contradiction to the incomparability of $N(a)$ and $N(a')$.

\noindent
{\bf Case 2.} $\iota_B(a) < \infty$, i.e., $a \in B$. Then $N(a)=\{\alpha_i,\dots,\alpha_r,\beta_{\iota_B(a)},\dots,\beta_s\} \supseteq N(a')$, a contradiction.
\qed

\medskip

From Lemma \ref{lm5} it follows that $r=s$ and that the elements of $X$ can be numbered in such a way that $A_i=\{x_1,\dots,x_i\}$, $i=1,\dots,r+1$.

\begin{lemma}
\label{lm6}
We have $B_i=\{x_{r+1},x_r,\dots,x_{r-i+2}\}$, $i=1,\dots,r+1$.
\end{lemma}

\noindent
{\em Proof.} Assume the contrary and choose $i$ maximal such that $B_i \neq \{x_{r+1},x_r,$ $\dots,x_{r-i+2}\}$. Then $i \le r$ since $B_{r+1}=A_{r+1}=X$. By the maximality of $i$ we have $B_{i+1}= \{x_{r+1},x_r,$ $\dots,x_{r-i+1}\}$ and by Lemma \ref{lm5}, $B_i=\{x_{r+1},x_r,$ $\dots,x_{r-i+1}\}\setminus \{x_j\}$ for some $j \in \{r-i+2,\dots,r+1\}$.
We have $j \neq r+1$ since otherwise $B_i \subseteq A_r$. Since $x_j\notin B_i$, the element $x_j$ is not contained in $B_1,\dots,B_i$. Thus $N(x_j)=\{\alpha_j,\dots,\alpha_r,\beta_r,\dots,\beta_{i+1}\} \subseteq \{\alpha_{r-i+1},\dots,\alpha_r,\beta_r,\dots,\beta_{i+1},\beta_i\} \subseteq N(x_{r-i+1})$, a contradiction.
\qed

\medskip

Now the equalities $A_i=\{x_1,\dots,x_i\}$ and $B_i=\{x_{r+1},x_r,\dots,x_{r-i+2}\}$, $i=1,\dots,r$, imply
\begin{align*}
N(x_1)&=\{\alpha_1,\dots,\alpha_r\},\\
N(x_i)&=\{\alpha_i,\dots,\alpha_r,\beta_r,\dots,\beta_{r-i+2}\}, i=2,\dots,r,\\
N(x_{r+1})&=\{\beta_r,\dots,\beta_1\}.
\end{align*}
Thus $G \sim B_r$. But since $G$ is $k$-critical, $G \sim B_k$ follows which finally shows Theorem \ref{ACBDilwkchar}.
\qed

\begin{corollary}\label{ACBgrDilwkcor}
The Dilworth number of an ACB graph $B=(X,Y,E)$ is at most $k$ if and only if $B$ is $B_{k+1}$-free.
\end{corollary}

Thus, Theorem \ref{ACBDilwkchar} yields a characterization by forbidden induced subgraphs both of the family
of ACB graphs with bipartite Dilworth number at most $k$, and of the split graph counterpart with Dilworth number at most $k$.
Since by Proposition \ref{Bk-1subBk}, $B_i$ is an induced subgraph of $B_k$ for each $i \in \{2,\dots,k\}$, this defines a hierarchy of properly included ACB graphs, with $B_k$ as separating example between
the class of Dilworth number $k$ and the class of Dilworth number $k-1$.
The corresponding hierarchy of split graphs is easily derived.

\section{Conclusion}\label{Sect:Algo}

In this paper, we have studied ACB graphs with respect to their Dilworth number. One of our open questions is the recognition complexity of ACB graphs.
Chordal bipartite graphs can be recognized in $O(\min(n\log n, n^2))$ \cite{Lubiw1987,PaiTar1987,Spinr1993}, by computing a doubly lexical ordering of the
bipartite adjacency matrix of the (bipartite) graph $G$, which is $\Gamma$-free if and only if $G$ is chordal bipartite (a $\Gamma$ is a submatrix with three 1's, and one 0 in the bottom-right corner). See \cite{Spinr2003} for a detailed discussion of this aspect. A linear-time recognition for chordal bipartite graphs remains a long-standing open question.

ACB graphs can be recognized in $O(n^2)$ time checking whether both the graph and its mirror are chordal bipartite. In Subsection \ref{relworkDilw2bip}, we discussed how to recognize whether a bipartite graph has $X$-Dilworth number ($Y$-Dil\-worth number, respectively) more than 2 in linear time.
An alternative linear-time approach for this would be to check whether $split_X(B)$ is an interval graph.
\par
\begin{figure}

\includegraphics[height=2cm]{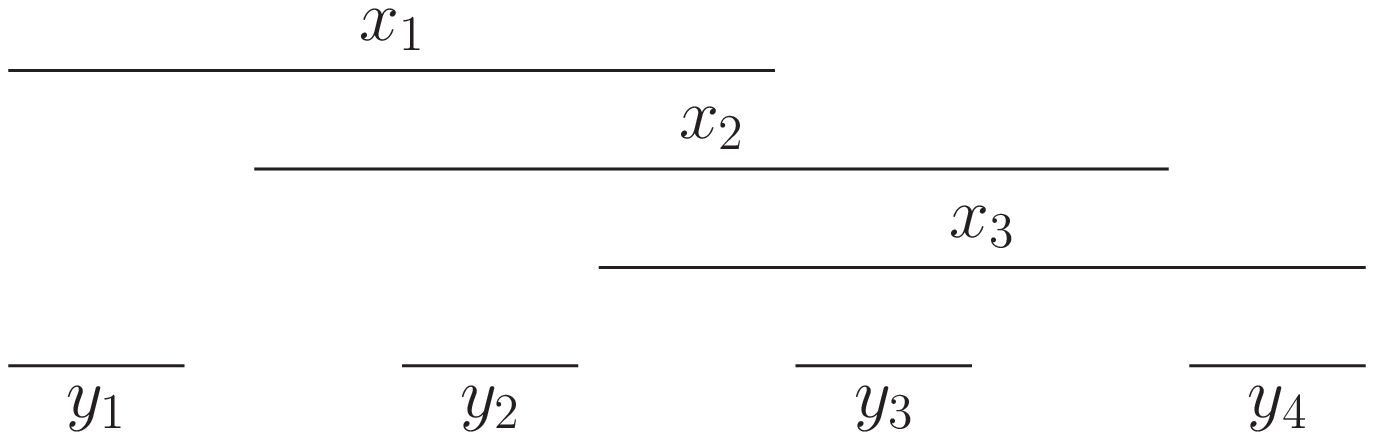}\hspace{1cm}
\begin{tabular}{l|l|l|l|l|}
& $y_1$ & $y_2$ & $y_3$ & $y_4$\\
\hline
$x_1$ & 1 & 1 & 0 & 0\\
\hline
$x_2$ & 0 & 1 & 1 & 0\\
\hline
$x_3$ & 0 & 0 & 1 & 1\\
\hline
\end{tabular}\\
\hspace*{1mm}\includegraphics[height=.6cm]{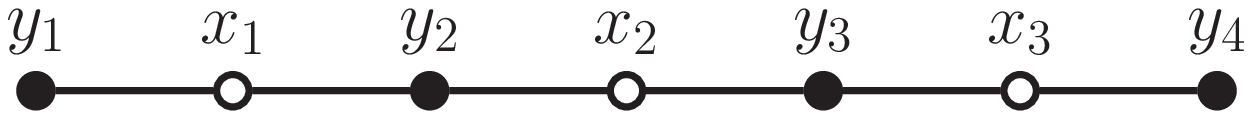}

\caption{Bipartite graph $B= Y$-$P_7$, an interval representation of $split_X(B)$, and its corresponding $\Gamma$-free matrix.\label{fig:gamma}}
\end{figure}

Let us remark that, given this interval realization, we can construct in linear time a $\Gamma$-free matrix of $B$ as follows:
order the vertices of $Y$ as they appear from left to right in the interval representation of  $split_X(B)$;
order the vertices of $X$ using the left endpoints of their intervals, and order these from left to right.
Figure \ref{fig:gamma} gives an example for the $P_7$, which has $X$-Dilworth number 3 and $Y$-Dil\-worth number 2.

This result is noteworthy, as to the best of our knowledge the chordal bipartite graphs for which there is a linear-time algorithm to produce a
$\Gamma$-free matrix are the (trivial) chain graphs.

\begin{corollary}\label{linrec3K2C6P7freebip}
It can be recognized in linear time if a given bipartite graph is $(3K_2,C_6,P_7)$-free, or if it is $(3K_2,C_6,X$-$P_7)$-free
\end{corollary}

Another open question is the complexity of computing the Dilworth number of an ACB graph:
Given a $\Gamma$-free matrix for a (chordal) bipartite graph, the neighborhood order can be computed in linear time \cite{Spinr2003}.
This order in turn yields the Dilworth number.
Thus as discussed above, if the $X$- or $Y$- Dilworth number is 2, the Dilworth number can be determined in linear time; if both the
$X$- and the $Y$- Dilworth numbers are $>2$, $O(n^2)$ time is required to compute the Dilworth number.

\medskip

We conclude this paper with the following open questions:
\begin{enumerate}
\item Can one recognize ACB graphs in linear time?
\item Determine all $(k,l)$-critical ACB graphs $G=(X,Y,E)$, i.e., all ACB graphs for which $\nabla_G(X)=k$, $\nabla_G(Y)=l$ and $\nabla_{G-v}(X)+\nabla_{G-v}(Y) < k+l$ for all $v \in X \cup Y$.
\end{enumerate}

\bibliographystyle{plain}

\end{document}